\documentclass[reprint,a4paper,aps,pra,amsmath,amsfonts,showpacs,floatfix,nofootinbib]{revtex4-2}
\usepackage{times}
\usepackage{amsmath}
\usepackage{times}
\usepackage{dcolumn}                 
\usepackage{graphicx}
\usepackage{xcolor}

\newcommand*{\centt}[1]{\multicolumn{1}{c}{#1}}
\newcommand*{\cent}[1]{\multicolumn{1}{c}{$#1$}}
\newcolumntype{x}[1]{D{.}{.}{#1}}

\newcommand{\header}[1]{\paragraph*{#1.---\hspace{-2ex}}}
\newcommand{\etal}{{\em et al.{}}}

\begin{document}
\title{Hyperfine Structure of the First Rotational Level in H$_2$, D$_2$ and HD Molecules
       and the Deuteron Quadrupole Moment}

\author{Mariusz Puchalski}
\affiliation{Faculty of Chemistry, Adam Mickiewicz University, Uniwersytetu Pozna{\'n}skiego 8, 61-614 Pozna{\'n}, Poland}

\author{Jacek Komasa}
\affiliation{Faculty of Chemistry, Adam Mickiewicz University, Uniwersytetu Pozna{\'n}skiego 8, 61-614 Pozna{\'n}, Poland}

\author{Krzysztof Pachucki}
\affiliation{Faculty of Physics, University of Warsaw, Pasteura 5, 02-093 Warsaw, Poland}

\date{\today}

\begin{abstract}
We perform the four-body calculation of the hyperfine structure in the first rotational state $J=1$  
of the H$_2$, D$_2$, and HD molecules and determine the accurate value for the deuteron electric
quadrupole moment $Q_d = 0.285\,699(15)(18)$ fm$^2$ in significant disagreement with former 
spectroscopic determinations. Our results for the hyperfine parameters agree very well with 
the currently most accurate molecular-beam magnetic resonance measurement performed several 
decades ago by N.F. Ramsey and coworkers. They also indicate the significance of previously neglected
nonadiabatic effects. Moreover, a very good agreement with the recent calculation 
of $Q_d$ based on the chiral effective field theory, although much less accurate,
indicates the importance of the spin dependence of nucleon interactions in the accurate description of nuclei. 
\end{abstract}

\maketitle

Precise atomic and molecular spectroscopy provides information on the nuclear electromagnetic moments 
important for testing theories of nuclear interactions \cite{Lev:99, Gilman:02, Ericson:83, Epelbaum:09, Filin:20b} or even for searching for new physics \cite{Safronowa:18}. This, however, requires a thorough understanding 
of the variety of interparticle interactions in atoms and molecules. For instance, the proton mean square 
charge radius has been extracted with unprecedented accuracy from the muonic hydrogen Lamb shift \cite{Pohl:10}
only after careful analysis of all the important quantum electrodynamical effects. A similar determination 
has recently been performed for the alpha particle from the corresponding measurement in muonic helium
\cite{Pohl:20}. Regarding nuclear magnetic moments, the direct measurement in a Penning trap was performed 
only for the proton \cite{Sturm:14}, while magnetic moments of heavier stable nuclei have been determined 
by the nuclear magnetic resonance or the atomic hyperfine splitting measurements. 
Accordingly, the currently most accurate magnetic moments of deuteron and triton were determined by combining 
the nuclear magnetic resonance measurements with precise calculations of the molecular shielding factor \cite{PKP:15}. 
Concerning the determination of nuclear magnetic moments from the hyperfine splitting, their accurate calculation 
is particularly difficult due to a large contribution from the not-well-known spin-dependent nuclear structure. 
The widely accepted Bohr-Weisskopf correction only partially accounts for the nuclear effects \cite{Bohr:50}.  
A clear indication of this problem is a strong and still unexplained discrepancy for the Zemach radius 
of $^{6}$Li between the nuclear model value \cite{Yerokhin:08} and the result based on the spectroscopic 
data of the lithium atom \cite{Puchalski:13}, which has recently been confirmed by independent measurements 
and calculations in the Li$^+$ ion \cite{Guan:20, Qi:20}.

In this work we investigate the electric quadrupole moment $Q_d$ of the deuteron on the basis 
of the hyperfine splitting in HD and D$_2$ molecules. The total electron spin of such a two-electron
system is zero, and the strengths of all couplings among nuclear spins and the rotational angular
momentum are of the same order of magnitude. Therefore, $Q_d$ can be extracted from the molecular
hyperfine splitting with an accuracy that is limited only by the measurement uncertainty provided 
that sufficiently accurate theoretical calculations with all significant contributions are available.

The recent determinations of $Q_d = 0.285\,98(3)$ fm$^2$  \cite{Jozwiak:20} and 
$Q_d = 0.285\,783(30)$ fm$^2$ \cite{Pavanello:10} considered to be the most accurate value to date
\cite{Pyykko:18}, neglect or underestimate nonadiabatic effects, i.e. the effects beyond 
the commonly employed Born-Oppenheimer (BO) approximation. Indeed, these results disagree with 
the recommended value reported in this work (see Eq.~(\ref{EQdrecom})) obtained in the nonadiabatic approach, i.e. without 
separation of nuclear and electronic motions.

In the following we describe shortly the theory of the molecular hyperfine splitting, 
its accurate calculations with nonadiabatic wave functions, and the determination 
of $Q_d$ from the measurements by Ramsey and coworkers.
Henceforth, we use the original notation by Ramsey \cite{Ramsey:56,Ramsey:57}.

\header{Hyperfine Hamiltonian}
There are three angular momenta in the ground electronic state of the heteronuclear HD molecule,
which all couple to each other---the proton spin $\vec{I}_p$, that of the deuteron $\vec{I}_d$, 
and the rotational angular momentum $\vec{J}$. The effective Hamiltonian describing these 
interactions reads
\begin{align}
  H_{\rm hfs} =&\ -c_p\,\vec I_p\cdot\vec J - c_d\,\vec I_d\cdot\vec J  \nonumber \\ &
  + \frac{5\,d_1}{(2\,J-1)(2\,J+3)}\,\biggl[
      \frac{3}{2}\,(\vec I_p\cdot\vec J)\, (\vec I_d\cdot\vec J) \nonumber  \\ &
 \hspace{1.5cm}  +\frac{3}{2}\,(\vec I_d\cdot\vec J)\, (\vec I_p\cdot\vec J)
      -(\vec I_p\cdot\vec I_d)\,\vec{J}^{\;2}\biggr]
\nonumber \\ &
 + \frac{5\,d_2}{(2\,J-1)(2\,J+3)}\,\biggl[
      3\,(\vec I_d\cdot\vec J)^2 \nonumber  \\ &
 \hspace{3cm} +\frac{3}{2}\,(\vec I_d\cdot\vec J)
      -\vec I_d^{\;2}\,\vec J^{\;2}\biggr]\,. 
\label{01}
\end{align}
The above coefficients $c_p$, $c_d$, $d_1$, and $d_2$ are related, respectively, 
to the interactions between the following: the proton spin and molecular rotation, 
the deuteron spin and rotation, the proton and deuteron spins, and the electric
quadrupole moment of the deuteron with the electric field gradient \cite{brown_carrington_2003}.
For homonuclear molecules with odd $\vec J$ the effective Hamiltonian takes a simplified form
\begin{align}
  H_{\rm hfs} =&\ -c\,\vec I\cdot\vec J
+ \frac{5\,d}{(2\,J-1)(2\,J+3)} \nonumber  \\ & \times\biggl[
      3\,(\vec I\cdot\vec J)^2
      +\frac{3}{2}\,(\vec I\cdot\vec J)
      -\vec I^{\;2}\,\vec J^{\;2}\biggr]\,,
\label{02}
\end{align}
where the total nuclear spin $\vec{I}=\vec I_A+\vec I_B$ is equal to 1, and
$c=c_p$, $d = d_1/2$ for H$_2$, and $c=c_d$, $d=d_1-d_2$ for D$_2$.
We will consider all these constants for the first rotational level $J=1$ and $v=0$
and present a short derivation of corresponding formulas followed by numerical calculations
using explicitly correlated wave functions. Both the derivation and the calculations 
are performed in the nonadiabatic regime.

The general spin-orbit Hamiltonian is of the form
\begin{align}
\delta H =& \sum_{\alpha,\beta}\frac{e_\alpha\,e_\beta}{4\,\pi}\,\frac{1}{2\,r_{\alpha\beta}^3}
\biggl[\frac{g_\alpha}{m_\alpha\,m_\beta}\,\vec I_\alpha\cdot\vec r_{\alpha\beta}\times\vec p_\beta
\nonumber \\ &\ 
  -\frac{(g_\alpha-1)}{m_\alpha^2}\,\vec I_\alpha\cdot\vec r_{\alpha\beta}\times \vec p_\alpha \biggr],
\label{03}
\end{align}
where the indices $\alpha$ and $\beta$ are for electrons and nuclei. 
The gyromagnetic factors
\begin{align}
  g_p &= \frac{\mu_p}{\mu_N\,I_p} =5.585\,695\ldots \label{04}\\
  g_d &= \frac{\mu_d}{\mu_N\,I_d}\,\frac{m_d}{m_p} = 1.714\,025\ldots\label{05}
\end{align}
are related to the magnetic moment of the proton $\mu_p = 2.792\,847\,344\,63(82)\,\mu_N$
and the deuteron $\mu_d = 0.857\,438\,2338(22)\,\mu_N$, respectively \cite{CODATA:18}.
The $g$ factor is a dimensionless quantity convenient for use in the magnetic moment formulas.
In particular, the coupling of the nuclear spin $\vec I_A$ 
to the molecular rotation, using Eq. (\ref{03}), is
\begin{align}
\delta_A H &= \vec I_A\cdot\vec Q_A\label{06}\\
\vec Q_A &= -\sum_{b}\frac{\alpha}{2\,r_{A b}^3}\!
  \biggl[\frac{g_A}{m_A\,m_e}\,\vec r_{A b}\times\vec p_b-
       \frac{(g_A-1)}{m_A^2}\,\vec r_{A b}\times \vec p_A \biggr]
\nonumber \\ &\quad
+\frac{\alpha}{2\,r_{A B}^3}
\biggl[\frac{g_A}{m_A\,m_B}\,\vec r_{A B}\times\vec p_B-
       \frac{(g_A-1)}{m_A^2}\,\vec r_{A B}\times \vec p_A \biggr] \label{07}
\end{align}
and the spin-rotation coefficient is thus
\begin{align}
  c_A = \frac{1}{2}\,\mathrm{i}\,\epsilon^{ijk}\,\langle\phi^i|Q_A^j|\phi^k\rangle\,, \label{08}
\end{align}  
where $\phi^i$ is the wave function for the first rotational state
with normalization $\langle\phi^i|\phi^i\rangle = 1$.

The nuclear spin-spin direct interaction can be effectively written as
\begin{align}
  \delta H =&\ I_A^i\,I_B^j\,Q_{AB}^{ij} \label{09}\\
Q_{AB}^{ij} =&\ \frac{g_A\,g_B}{4\,m_A\,m_B}\,\frac{\alpha}{r_{AB}^3}\,\biggl(\delta^{ij}-3\,\frac{r_{AB}^i\,r_{AB}^j}{r_{AB}^2}\biggr)
\label{10}
\end{align}
and the corresponding $d_1$ coefficient is
\begin{align}
  d_1 = -\frac{1}{5}\,\langle\phi^i| Q_{AB}^{ij} |\phi^j\rangle \label{Ed1}
\end{align}
The omitted part of the spin-spin interaction, proportional to $\delta^3(r_{AB})$,
is negligibly small.

The interaction of a particle with the charge $e$,  possessing the electric quadrupole moment $Q^{ij}$ with 
the gradient of the electric field, is given by
\begin{align}
  \delta H =&\ -\frac{e}{6}\,Q^{ij}\,\partial_jE^i\,.\label{12}
\end{align}
For a particle with a definite spin $I\geq 1$, the $Q^{ij}$, as a traceless and symmetric tensor, 
can be expressed in terms of a single scalar electric quadrupole moment $Q$ defined by 
\begin{align}
  Q^{ij} =&\ \frac{Q}{I\,(2I-1)}\,\biggl(\frac{3}{2}\,I^i\,I^j + \frac{3}{2} I^j\,I^i -\delta^{ij}\,\vec I^2\biggr).\label{13}
\end{align}
Referring to Eq.~(\ref{01}), the Ramsey's constant $d_2$ becomes (in atomic units)
\begin{align}
  d_2&=-\alpha^2\,\frac{Q\,q}{10\,\lambdabar^2},\label{Ed2}
\intertext{where}
  q&=\left\langle\phi^i\left|\frac{\partial^2 V}{\partial r_A^i\,\partial r_A^j}\,
  -\frac{\delta^{ij}}{3}\frac{\partial^2 V}{\partial r_A^k\,\partial r_A^k}\right|\phi^j\right\rangle\label{Eq}
\end{align}
is the electric field gradient at the nucleus $A$, $V$ is the Coulomb potential of Eq.~(\ref{EV}), 
and $\lambdabar$ is the reduced Compton wavelength of the electron.

\header{Numerical calculations}
The nonadiabatic wave function $\Psi$ is obtained from the variational principle
with the following nonrelativistic Hamiltonian  for the hydrogen molecule
\begin{equation} \label{16}
H = T + V\,,
\end{equation}
where (in atomic units)
\begin{eqnarray}
T &=& \frac{ \vec p_A^{\,2}}{2\,m_A} + \frac{ \vec p_B^{\,2}}{2\,m_B} +  \frac{\vec p_1^{\,2}}{2} + \frac{\vec p_2^{\,2}}{2} \,, \label{17}\\
V &=& \frac{1}{r_{AB}} -\frac{1}{r_{A1}} - \frac{1}{r_{A2}}-\frac{1}{r_{B1}} - \frac{1}{r_{B2}} +\frac{1}{r_{12}} \,.\label{EV}
\end{eqnarray}
Here, indices $A$, $B$ and $1$, $2$ denote nuclei and electrons, respectively. 
The nuclear masses are those currently recommended by CODATA \citep{CODATA:18}.
The wave function $\Psi$ depends on four particle coordinates 
$\Psi = \Psi(\vec r_A, \vec r_B, \vec r_1, \vec r_2)$.
In the center of mass frame the total momentum vanishes
$\vec{p}_A + \vec p_B + \vec p_1 + \vec p_2 = 0 $,
and thus we may assume that the wave function $\Psi$ depends only on the coordinate differences.

In the variational approach the wave function is represented as a linear combination
\begin{equation}
  \Psi = \sum_k^N c_k\, \psi_k(\vec r_A, \vec r_B, \vec r_1, \vec r_2)\,,\label{19}
\end{equation}
of properly symmetrized basis functions
\begin{equation}
 \psi_k = (1 \pm P_{A\leftrightarrow B})\,(1+P_{1\leftrightarrow 2})\,\phi_k (\vec r_A, \vec r_B, \vec r_1, \vec r_2), \label{Epsik}
 \end{equation}
where $P_{i\leftrightarrow j}$ is the particle exchange operator. In the $J=1$ state
of H$_2$ and D$_2$ the wave function is antisymmetric with respect to the exchange 
of nuclear spatial variables and symmetric in electronic spatial variables,
whereas in the heteronuclear HD molecule,
only electronic symmetry is imposed, and both (nuclear) symmetric and antisymmetric 
basis functions are employed. For $J=1$ the functions $\phi_k$ in Eq. (\ref{Epsik}) 
are the nonadiabatic explicitly correlated Gaussians (naECG) of the form
\begin{align}
  \phi^i_k =&\ r^i\,r_{AB}^n \label{21} \\ \times &
  e^{-a_{k,1}\, r^2_{AB}-a_{k,2}\, r^2_{A1}-a_{k,3}\, r^2_{A2}-a_{k,4}\, r^2_{B1}-a_{k,5}\, r^2_{B2}-a_{k,6}\, r^2_{12}}, \nonumber
\end{align}
where $\vec{r}$ (or $r^i$) is the factor representing the angular momentum $J=1$, and is 
either $\vec{r}_{AB}$, $\vec{r}_{A1}$, $\vec{r}_{A2}$, $\vec{r}_{B1}$, $\vec{r}_{B2}$ or $\vec{r}_{12}$.
The nonlinear $a_k$ parameters are optimized individually for each basis function $\phi_k$.
The powers $n$ of the internuclear coordinate $r_{AB}$, needed to represent accurately 
the vibrational part of the wave function, are restricted to even integers 
and are generated randomly for each basis function from the log-normal distribution 
within the limited 0-80 range. Moreover, the number of basis functions with the particular prefactor 
$\vec {r}$\,$r_{AB}^n$ is subject to additional discrete optimization. As a result, the nonrelativistic 
energy reaches an accuracy of about $10^{-11}$ (see Table~\ref{Tconv}).

\renewcommand{\arraystretch}{1.0}
\begin{table*}[!htb] 
\caption{Convergence of the nonrelativistic energy $E$ (in a.u.), hyperfine splitting 
parameters (in kHz), and the electric field gradient $q$ (in a.u.) calculated using naECG 
wave functions for the first rotational level $(v,J) = (0,1)$. 
The extrapolated nonadiabatic results (NA) are compared to the Born-Oppenheimer 
(BO) values \cite{Komasa:20} and with the results of measurements by Ramsey 
\etal~\cite{Harrick:53,Quinn:58,Code:71}. 
The NA values do not include here the uncertainties due to omitted relativistic 
and QED effects, which can be estimated by the relative factor of $\alpha^2\approx5\times10^{-5}$, 
where $\alpha$ is the fine structure constant.}
\label{Tconv}
\begin{ruledtabular}
\begin{tabular}{cx{3.18}@{\extracolsep{\fill}}*{2}{x{3.6}}*{2}{x{5.10}}}
\multicolumn{6}{c}{H$_2$} \\
 Basis   &\cent{E}& \cent{c_p}  & & \cent{d=d_1/2} & \\[1ex] 
 \hline
512  & -1.163\,485\,167\,695         & 112.393\,16 & & 57.643\,938\,904\,75  \\
1024 & -1.163\,485\,172\,061         & 113.889\,80 & & 57.643\,937\,929\,23  \\
1536 & -1.163\,485\,172\,209         & 113.904\,33 & & 57.643\,937\,899\,30  \\
2048 & -1.163\,485\,172\,287         & 113.911\,88 & & 57.643\,937\,895\,69  \\
 NA  & -1.163\,485\,172\,314\,0(1)^a & 113.920(8)  & & 57.643\,937\,891(6)   \\[2ex]
 BO, \cite{Komasa:20} &              & 114.00(12)  & & 57.69(6)              \\
NA $-$ BO&                           &  -0.08(12)  & & -0.05(6)              \\
Measured, \cite{Harrick:53} &        & 113.904(30) & & 57.671(24)            \\[2ex]
\multicolumn{6}{c}{HD} \\
 Basis   &\cent{E}& \cent{c_p}  & \cent{c_d} & \cent{d_1} & \cent{q} \\[1ex] 
 \hline
512  &-1.165\,065\,367\,519         & 83.740\,0  & 13.281\,98 & 17.761\,872\,179\,98 & 0.334\,510\,892 \\ 
1024 &-1.165\,065\,376\,045         & 85.446\,7  & 13.146\,42 & 17.761\,872\,391\,72 & 0.334\,493\,539 \\ 
1536 &-1.165\,065\,376\,735         & 85.550\,6  & 13.135\,55 & 17.761\,872\,423\,13 & 0.334\,492\,630 \\
2048 &-1.165\,065\,376\,858         & 85.598\,6  & 13.126\,23 & 17.761\,872\,410\,33 & 0.334\,491\,813 \\
NA   &-1.165\,065\,376\,941\,65(3)^a& 85.63(4)   & 13.117(9)  & 17.761\,872\,414(13) & 0.334\,491\,0(8)\\[2ex]
BO, \cite{Komasa:20}      &         & 85.675(60) & 13.132(9)  & 17.773(12)           & 0.334\,7(3) \\
NA $-$ BO&                          & -0.04(7)   & -0.014(14) & -0.012(12)           &-0.000\,2(3) \\
Measured, \cite{Quinn:58} &         & 85.600(18) & 13.122(11) & 17.761(12)           & \\[2ex]
\multicolumn{6}{c}{D$_2$} \\
 Basis   &\cent{E}&  & \cent{c_d} & \cent{d_1} & \cent{q} \\[1ex] 
 \hline
512  &-1.166\,896\,428\,705          & & 8.723\,72     & 2.737\,626\,131\,22 & 0.335\,240\,662 \\ 
1024 &-1.166\,896\,432\,071          & & 8.763\,77     & 2.737\,626\,043\,01 & 0.335\,233\,684 \\ 
1536 &-1.166\,896\,432\,230          & & 8.765\,41     & 2.737\,626\,038\,04 & 0.335\,232\,171 \\
2048 &-1.166\,896\,432\,323          & & 8.766\,20     & 2.737\,626\,037\,75     & 0.335\,231\,377 \\
NA   &-1.166\,896\,432\,359\,76(4)^a & & 8.767\,4(10)  & 2.737\,626\,037\,4(12)  & 0.335\,230\,7(7)\\[2ex]
BO, \cite{Komasa:20}     &           & & 8.770(5)      & 2.739(2)            & 0.335\,35(18)   \\
NA $-$ BO&                           & &-0.003(5)      &-0.002(2)            &-0.000\,12(18)   \\
Measured, \cite{Code:71} &           & & 8.768(3)      &                     &                 \\
\end{tabular}
\end{ruledtabular}
\flushleft
$^a$ This is a reference energy obtained from explicitly correlated exponential functions 
\cite{PK:18a,PK:18b,PK:19}
\end{table*} 
 
\header{Hyperfine parameters}
The hyperfine parameters for the hydrogen molecule isotopologues obtained with the above wave function
are presented in Table~\ref{Tconv}. The numerical convergence for the spin-orbit couplings 
$c_p$ and $c_d$ is relatively slow, and the resulting numerical uncertainties are not negligible. 
Most importantly, their difference from BO values fits within the uncertainties, 
which indicates that the estimation of the magnitude of nonadiabatic effects by the ratio 
of the electron mass to the nuclear reduced mass is correct.
Moreover, the nonadiabatic and BO values are, within uncertainties, in agreement with 
the Ramsey measurements.

In contrast, the numerical convergence of $d_1$ and $q$ parameters is very fast,
and the corresponding inaccuracy is negligible compared to the uncertainty due to 
unknown higher order relativistic and QED effects. Again, the difference 
with BO values is consistent with the estimate of nonadiabatic effects, represented 
as an inaccuracy of the BO values. We also note that the nonadiabatic $d_1$ for HD
agrees with Ramsey's measurements up to its uncertainty, while for H$_2$ it fits within
$1.2\,\sigma$. Regarding the $q$ parameter, the fast numerical convergence of the gradient of
the electric field enables six significant digits to be quoted. 
Our recommended nonadiabatic value of 
\begin{equation}
q=0.335\,230\,7(7) \text{ a.u. }
\end{equation}
obtained for D$_2$ 
will be used in the next paragraph for the determination of the deuteron quadrupole moment $Q_d$.
We note that this value differs by $0.000\,6$ a.u. from the $q=0.334\,66$ a.u obtained in
the pioneering nonadiabatic calculations by Bishop and Cheung \cite{Bishop:79}. 
This difference is relatively large and shows that the former results in \cite{Bishop:79} are not
accurate enough to draw definite conclusions about the magnitude of the nonadiabatic effects.

\begin{figure}[b]
\includegraphics[scale=0.21]{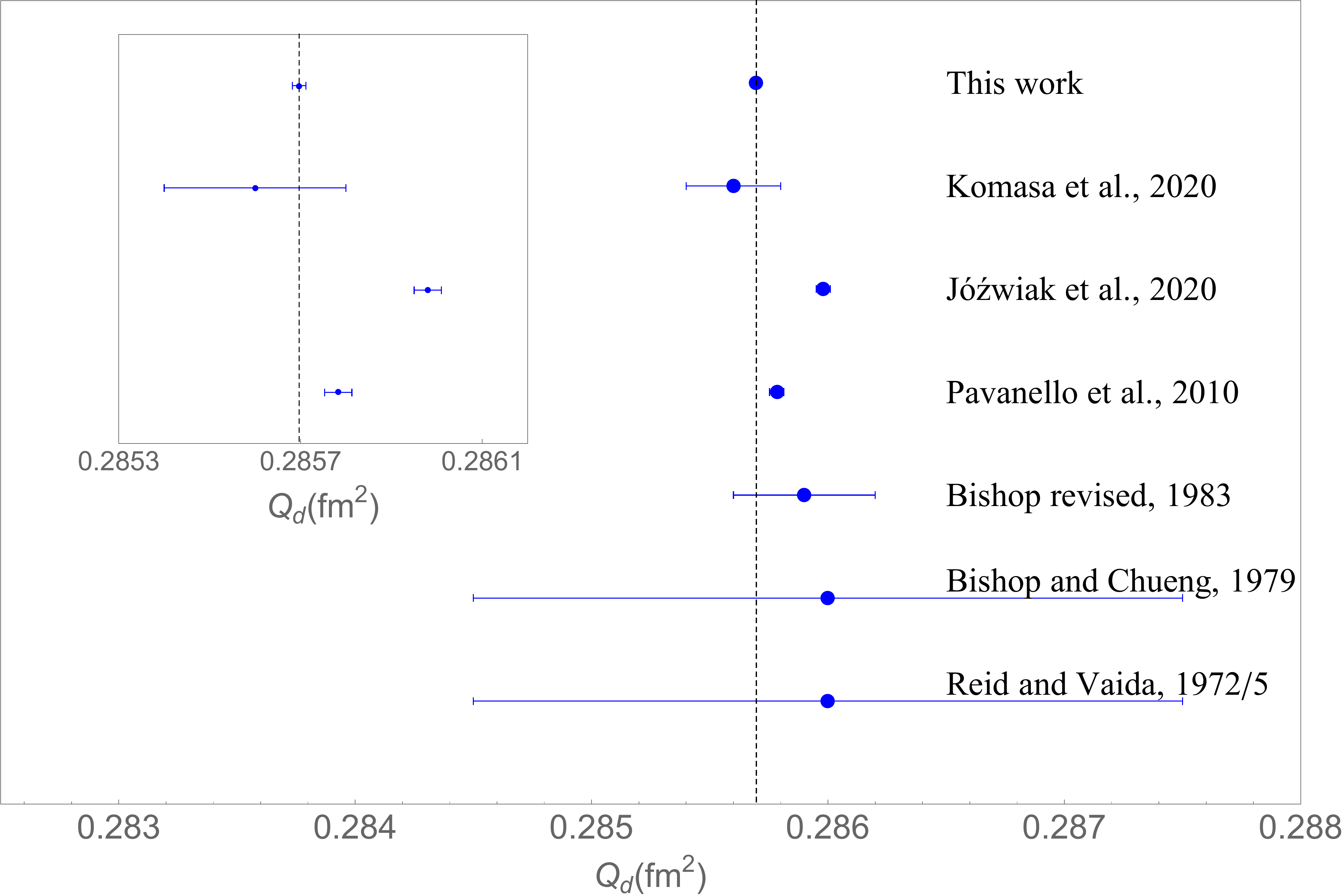}
\caption{Comparison of literature data for $Q_d$.}
\label{FQdFig}
\end{figure}

\header{The quadrupole moment of deuteron}
$Q_d$ can be determined most accurately from the coupling 
constant $d$ measured by the molecular-beam magnetic resonance method of Code and Ramsey \cite{Code:71}
for D$_2$ in the first rotational level. We obtain $d_2$ from the difference $d_2=d_1-d$
of calculated $d_1$ and measured $d$. Finally, we evaluate the quadrupole moment from
\begin{align}\label{EQd}
  Q_d &= -\frac{d_2}{2\,Ry\,c}\,\frac{10\,\lambdabar^2}{\alpha^2\,q}
\intertext{obtaining the recommended value of}
Q_d&=0.285\,699(15)(18) \text{ fm$^2$. } \label{EQdrecom}
\end{align}
The results of such calculations are summarized in Table~\ref{TQd}.
A comparison with literature data reported within the last fifty years is presented in Table~\ref{TQd} 
and Fig~\ref{FQdFig}. The first uncertainty assigned to our $Q_d$ is due to unknown higher order relativistic and QED corrections to $q$, which are estimated by a relative factor $\alpha^2$. The second uncertainty comes from the measurement of $d$, while numerical uncertainties are negligible. 
We should mention that the second order magnetic dipole interaction,
being also a kind of relativistic correction, leads to the pseudoquadrupole effect
estimated by Ramsey to be of about $10^{-5}$ \cite{Ramsey:53b} 
and this contribution is included in our $\alpha^2$ uncertainty.

This nonadiabatic $Q_d$ is in agreement with 
our BO value \cite{Komasa:20}, the relative uncertainty of which was estimated 
by the ratio of the electron mass to the nuclear reduced mass. It is in significant 
disagreement, though,  with the recent most accurate determination by Pavanello \etal~\cite{Pavanello:10}
and J\'o\'zwiak \etal~\cite{Jozwiak:20} (see Table~\ref{TQd} and Fig~\ref{FQdFig}). 
It is in agreement, however, with the revised result by Bishop, quoted in 1983 by Ericson 
\etal~\cite{Ericson:83}, 
which served for a long time as a reference value for $Q_d$. A comparison of our result with recent 
literature data indicates the significance of nonadiabatic effects and also draws attention 
to the need for the correct use of the BO potential, as the above mentioned results 
\cite{Pavanello:10,Jozwiak:20,Komasa:20} 
differ from each other due to different potentials used to average the $q$ parameter.
To verify our result, we used the obtained value of $Q_d$ to evaluate
the $d_2$ parameter for the HD molecule in a $J=1$ state, and we achieved a perfect agreement with $d_2$
as measured by Ramsey \etal~\cite{Quinn:58} (see Table~\ref{TQd}). 

\renewcommand{\arraystretch}{1.0}
\begin{table}[!bh] 
\caption{Determination of the deuteron quadrupole moment (in fm$^2$) from D$_2$ hyperfine parameters 
 and its verification for HD.}
 \label{TQd}
\begin{ruledtabular}
\begin{tabular*}{0.65\textwidth}{c@{\extracolsep{\fill}}x{3.18}l}
Quantity        &                    & Source  \\ 
                & \centt{D$_2$}      &         \\ 
\hline
$d$           & 25.241\,4(14)\text{ kHz}      & Code \& Ramsey, 1971 \cite{Code:71}  \\[1ex] 
$d_1$         &  2.737\,626\,037\,4(12)\text{ kHz}& This work, Eq.~(\ref{Ed1}) \\
$d_2$         &-22.503\,8(14)\text{ kHz}      & $d_2 = d_1-d$ \\
$q$           &  0.335\,230\,7(7)\text{ a.u.}  & This work, Eq.~(\ref{Eq}) \\[2ex]
$Q_d$         &  0.285\,699(15)(18)^a& This work, Eq.~(\ref{EQd})\\
              &  0.285\,4_{-17}^{+38\ b} & Filin \etal, 2020 \cite{Filin:20a} \\
              &  0.285\,98(3)      & J\'o\'zwiak \etal, 2020 \cite{Jozwiak:20} \\
              &  0.285\,6(2)       & Komasa \etal, 2020 \cite{Komasa:20} \\
              &  0.285\,783(30)    & Pavanello \etal, 2010 \cite{Pavanello:10} \\
              &  0.285\,90(30)     & Bishop revisited, 1983, cited in \cite{Ericson:83} \\
              &  0.286\,0(15)      & Bishop \& Cheung, 1979 \cite{Bishop:79} \\
              &  0.286\,0(15)      & Reid \& Vaida, 1972/5 \cite{Reid:72,Reid:75Errata} \\[1ex]
              & \centt{HD}         & \\ \hline
$q$           &  0.334\,491\,0(8)\text{ a.u.}  & This work, Eq.~(\ref{Eq}) \\
$d_2$         &-22.454\,2(14)^c\text{ kHz}& This work, Eq.~(\ref{Ed2}) \\
$d_2$         &-22.454(6)\text{ kHz}          & Ramsey \etal, 1958 \cite{Quinn:58} \\
Diff.         & -0.000(6)\text{ kHz}         \\
\end{tabular*}
\end{ruledtabular}
\flushleft
$^a$ The first uncertainty accounts for missing relativistic correction, the second one is
due to the experimental 
uncertainty of $d$.\\
$^b$ Calculated in the framework of $\chi$EFT. \\
$^c$ This uncertainty comes from the experimental uncertainty in $Q_d$.
\end{table} 

\header{Summary}
The deuteron quadrupole moment $Q_d$ is determined with the highest accuracy among all nuclei
in the periodic table \cite{Pyykko:18}. This accurate result can be used not only in precise 
atomic and molecular
structure calculations but also can serve as a benchmark for the nuclear structure theory.
Indeed, for a long time, $Q_d$ could not be reproduced by any modern potential model 
(see e.g.~\cite{Ericson:83} and references therein). It was not until very recently 
very recently Filin \etal~\cite{Filin:20a} reported $Q_d=0.285\,4_{-17}^{+38}$ fm$^2$ 
obtained from chiral effective field theory ($\chi$EFT), which is in very good agreement with our result.
This agreement opens the possibility of better understanding the spin-dependent 
nuclear structure effects in atomic spectra, particularly in muonic deuterium hyperfine splitting,
where significant discrepancies with measurement \cite{Pohl:16} have been reported 
\cite{Kalinowski:18}.

Apart from the deuteron quadrupole moment, by accounting for nonadiabatic effects,
we obtained all the other hyperfine constants in very good agreement with Ramsey's 
molecular-beam magnetic resonance measurements  (see Table I). 
However, current theory includes only the leading relativistic effects.
Because the inclusion of higher order relativistic and QED corrections is certainly within reach, 
more accurate measurements are desirable. This may open new windows
for high-precision tests of fundamental interactions on the molecular scale.

This research was supported by National Science Center (Poland) Grant No. 2016/23/B/ST4/01821
as well as by a computing grant from Pozna\'n Supercomputing and Networking Center and by PL-Grid Infrastructure.


\begin{thebibliography}{36}
\expandafter\ifx\csname natexlab\endcsname\relax\def\natexlab#1{#1}\fi
\expandafter\ifx\csname bibnamefont\endcsname\relax
  \def\bibnamefont#1{#1}\fi
\expandafter\ifx\csname bibfnamefont\endcsname\relax
  \def\bibfnamefont#1{#1}\fi
\expandafter\ifx\csname citenamefont\endcsname\relax
  \def\citenamefont#1{#1}\fi
\expandafter\ifx\csname url\endcsname\relax
  \def\url#1{\texttt{#1}}\fi
\expandafter\ifx\csname urlprefix\endcsname\relax\def\urlprefix{URL }\fi
\providecommand{\bibinfo}[2]{#2}
\providecommand{\eprint}[2][]{\url{#2}}

\bibitem[{\citenamefont{Lev et~al.}(1999)\citenamefont{Lev, Pace, and
  Salm\`e}}]{Lev:99}
\bibinfo{author}{\bibfnamefont{F.~M.} \bibnamefont{Lev}},
  \bibinfo{author}{\bibfnamefont{E.}~\bibnamefont{Pace}}, \bibnamefont{and}
  \bibinfo{author}{\bibfnamefont{G.}~\bibnamefont{Salm\`e}},
  \bibinfo{journal}{Phys. Rev. Lett.} \textbf{\bibinfo{volume}{83}},
  \bibinfo{pages}{5250} (\bibinfo{year}{1999}).

\bibitem[{\citenamefont{Gilman and Gross}(2002)}]{Gilman:02}
\bibinfo{author}{\bibfnamefont{R.}~\bibnamefont{Gilman}} \bibnamefont{and}
  \bibinfo{author}{\bibfnamefont{F.}~\bibnamefont{Gross}}, \bibinfo{journal}{J.
  Phys. G} \textbf{\bibinfo{volume}{28}}, \bibinfo{pages}{R37}
  (\bibinfo{year}{2002}).

\bibitem[{\citenamefont{Ericson and Rosa-Clot}(1983)}]{Ericson:83}
\bibinfo{author}{\bibfnamefont{T.}~\bibnamefont{Ericson}} \bibnamefont{and}
  \bibinfo{author}{\bibfnamefont{M.}~\bibnamefont{Rosa-Clot}},
  \bibinfo{journal}{Nucl. Phys. A} \textbf{\bibinfo{volume}{405}},
  \bibinfo{pages}{497 } (\bibinfo{year}{1983}).

\bibitem[{\citenamefont{Epelbaum et~al.}(2009)\citenamefont{Epelbaum, Hammer,
  and Mei\ss{}ner}}]{Epelbaum:09}
\bibinfo{author}{\bibfnamefont{E.}~\bibnamefont{Epelbaum}},
  \bibinfo{author}{\bibfnamefont{H.-W.} \bibnamefont{Hammer}},
  \bibnamefont{and} \bibinfo{author}{\bibfnamefont{Ulf-G.}
  \bibnamefont{Mei\ss{}ner}}, \bibinfo{journal}{Rev. Mod. Phys.}
  \textbf{\bibinfo{volume}{81}}, \bibinfo{pages}{1773} (\bibinfo{year}{2009}).

\bibitem[{\citenamefont{Filin et~al.}(2020{\natexlab{a}})\citenamefont{Filin,
  Baru, Epelbaum, Krebs, M\"oller, and Reinert}}]{Filin:20b}
\bibinfo{author}{\bibfnamefont{A.~A.} \bibnamefont{Filin}},
  \bibinfo{author}{\bibfnamefont{V.}~\bibnamefont{Baru}},
  \bibinfo{author}{\bibfnamefont{E.}~\bibnamefont{Epelbaum}},
  \bibinfo{author}{\bibfnamefont{H.}~\bibnamefont{Krebs}},
  \bibinfo{author}{\bibfnamefont{D.}~\bibnamefont{M\"oller}}, \bibnamefont{and}
  \bibinfo{author}{\bibfnamefont{P.}~\bibnamefont{Reinert}},
  \bibinfo{journal}{Phys. Rev. Lett.} \textbf{\bibinfo{volume}{124}},
  \bibinfo{pages}{082501} (\bibinfo{year}{2020}{\natexlab{a}}).

\bibitem[{\citenamefont{Safronova et~al.}(2018)\citenamefont{Safronova, Budker,
  DeMille, Kimball, Derevianko, and Clark}}]{Safronowa:18}
\bibinfo{author}{\bibfnamefont{M.~S.} \bibnamefont{Safronova}},
  \bibinfo{author}{\bibfnamefont{D.}~\bibnamefont{Budker}},
  \bibinfo{author}{\bibfnamefont{D.}~\bibnamefont{DeMille}},
  \bibinfo{author}{\bibfnamefont{Derek~F.~Jackson} \bibnamefont{Kimball}},
  \bibinfo{author}{\bibfnamefont{A.}~\bibnamefont{Derevianko}},
  \bibnamefont{and} \bibinfo{author}{\bibfnamefont{C.~W.} \bibnamefont{Clark}},
  \bibinfo{journal}{Rev. Mod. Phys.} \textbf{\bibinfo{volume}{90}},
  \bibinfo{pages}{025008} (\bibinfo{year}{2018}).

\bibitem[{\citenamefont{Pohl et~al.}(2010)\citenamefont{Pohl, Antognini, Nez,
  Amaro, Biraben, Cardoso, Covita, Dax, Dhawan, Fernandes et~al.}}]{Pohl:10}
\bibinfo{author}{\bibfnamefont{R.}~\bibnamefont{Pohl}},
  \bibinfo{author}{\bibfnamefont{A.}~\bibnamefont{Antognini}},
  \bibinfo{author}{\bibfnamefont{F.}~\bibnamefont{Nez}},
  \bibinfo{author}{\bibfnamefont{F.~D.} \bibnamefont{Amaro}},
  \bibinfo{author}{\bibfnamefont{F.}~\bibnamefont{Biraben}},
  \bibinfo{author}{\bibfnamefont{J.~M.~R.} \bibnamefont{Cardoso}},
  \bibinfo{author}{\bibfnamefont{D.~S.} \bibnamefont{Covita}},
  \bibinfo{author}{\bibfnamefont{A.}~\bibnamefont{Dax}},
  \bibinfo{author}{\bibfnamefont{S.}~\bibnamefont{Dhawan}},
  \bibinfo{author}{\bibfnamefont{L.~M.~P.} \bibnamefont{Fernandes}},
  \bibnamefont{et~al.}, \bibinfo{journal}{Nature}
  \textbf{\bibinfo{volume}{466}}, \bibinfo{pages}{213} (\bibinfo{year}{2010}).

\bibitem[{\citenamefont{Pohl and coworkers}(private communication)}]{Pohl:20}
\bibinfo{author}{\bibfnamefont{R.}~\bibnamefont{Pohl}} \bibnamefont{and}
  \bibinfo{author}{\bibnamefont{coworkers}} (\bibinfo{year}{private
  communication}).

\bibitem[{\citenamefont{Sturm et~al.}(2014)\citenamefont{Sturm, Kohler,
  Zatorski, Wagner, Harman, Werth, Quint, Keitel, and Blaum}}]{Sturm:14}
\bibinfo{author}{\bibfnamefont{S.}~\bibnamefont{Sturm}},
  \bibinfo{author}{\bibfnamefont{F.}~\bibnamefont{Kohler}},
  \bibinfo{author}{\bibfnamefont{J.}~\bibnamefont{Zatorski}},
  \bibinfo{author}{\bibfnamefont{A.}~\bibnamefont{Wagner}},
  \bibinfo{author}{\bibfnamefont{Z.}~\bibnamefont{Harman}},
  \bibinfo{author}{\bibfnamefont{G.}~\bibnamefont{Werth}},
  \bibinfo{author}{\bibfnamefont{W.}~\bibnamefont{Quint}},
  \bibinfo{author}{\bibfnamefont{C.~H.} \bibnamefont{Keitel}},
  \bibnamefont{and} \bibinfo{author}{\bibfnamefont{K.}~\bibnamefont{Blaum}},
  \bibinfo{journal}{Nature} \textbf{\bibinfo{volume}{506}},
  \bibinfo{pages}{467} (\bibinfo{year}{2014}).

\bibitem[{\citenamefont{Puchalski et~al.}(2015)\citenamefont{Puchalski, Komasa,
  and Pachucki}}]{PKP:15}
\bibinfo{author}{\bibfnamefont{M.}~\bibnamefont{Puchalski}},
  \bibinfo{author}{\bibfnamefont{J.}~\bibnamefont{Komasa}}, \bibnamefont{and}
  \bibinfo{author}{\bibfnamefont{K.}~\bibnamefont{Pachucki}},
  \bibinfo{journal}{Phys. Rev. A} \textbf{\bibinfo{volume}{92}},
  \bibinfo{pages}{020501(R)} (\bibinfo{year}{2015}).

\bibitem[{\citenamefont{Bohr and Weisskopf}(1950)}]{Bohr:50}
\bibinfo{author}{\bibfnamefont{A.}~\bibnamefont{Bohr}} \bibnamefont{and}
  \bibinfo{author}{\bibfnamefont{V.~F.} \bibnamefont{Weisskopf}},
  \bibinfo{journal}{Phys. Rev.} \textbf{\bibinfo{volume}{77}},
  \bibinfo{pages}{94} (\bibinfo{year}{1950}).

\bibitem[{\citenamefont{Yerokhin}(2008)}]{Yerokhin:08}
\bibinfo{author}{\bibfnamefont{V.~A.} \bibnamefont{Yerokhin}},
  \bibinfo{journal}{Phys. Rev. A} \textbf{\bibinfo{volume}{78}},
  \bibinfo{pages}{012513} (\bibinfo{year}{2008}).

\bibitem[{\citenamefont{Puchalski and Pachucki}(2013)}]{Puchalski:13}
\bibinfo{author}{\bibfnamefont{M.}~\bibnamefont{Puchalski}} \bibnamefont{and}
  \bibinfo{author}{\bibfnamefont{K.}~\bibnamefont{Pachucki}},
  \bibinfo{journal}{Phys. Rev. Lett.} \textbf{\bibinfo{volume}{111}},
  \bibinfo{pages}{243001} (\bibinfo{year}{2013}).

\bibitem[{\citenamefont{Guan et~al.}(2020)\citenamefont{Guan, Chen, Qi, Liang,
  Sun, Zhou, Huang, Zhang, Zhong, Yan et~al.}}]{Guan:20}
\bibinfo{author}{\bibfnamefont{H.}~\bibnamefont{Guan}},
  \bibinfo{author}{\bibfnamefont{S.}~\bibnamefont{Chen}},
  \bibinfo{author}{\bibfnamefont{X.-Q.} \bibnamefont{Qi}},
  \bibinfo{author}{\bibfnamefont{S.}~\bibnamefont{Liang}},
  \bibinfo{author}{\bibfnamefont{W.}~\bibnamefont{Sun}},
  \bibinfo{author}{\bibfnamefont{P.}~\bibnamefont{Zhou}},
  \bibinfo{author}{\bibfnamefont{Y.}~\bibnamefont{Huang}},
  \bibinfo{author}{\bibfnamefont{P.-P.} \bibnamefont{Zhang}},
  \bibinfo{author}{\bibfnamefont{Z.-X.} \bibnamefont{Zhong}},
  \bibinfo{author}{\bibfnamefont{Z.-C.} \bibnamefont{Yan}},
  \bibnamefont{et~al.}, \bibinfo{journal}{Phys. Rev. A}
  \textbf{\bibinfo{volume}{102}}, \bibinfo{pages}{030801(R)}
  (\bibinfo{year}{2020}).

\bibitem{Qi:20} Xiao-Qiu Qi, Pei-Pei Zhang, Zong-Chao Yan, G.W.F.  Drake, Zhen-Xiang Zhong, Ting-Yun Shi, Shao-Long Chen, Yao Huang, Hua Guan, and Ke-Lin Gao, 
Phys. Rev. Lett. {\bf 125}, 183002 (2020).

\bibitem[{\citenamefont{J{\'o}{\'z}wiak
  et~al.}(2020)\citenamefont{J{\'o}{\'z}wiak, Cybulski, and
  Wcis{\l}o}}]{Jozwiak:20}
\bibinfo{author}{\bibfnamefont{H.}~\bibnamefont{J{\'o}{\'z}wiak}},
  \bibinfo{author}{\bibfnamefont{H.}~\bibnamefont{Cybulski}}, \bibnamefont{and}
  \bibinfo{author}{\bibfnamefont{P.}~\bibnamefont{Wcis{\l}o}},
  \bibinfo{journal}{J. Quant. Spectrosc. Radiat. Transfer}
  \textbf{\bibinfo{volume}{253}}, \bibinfo{pages}{107186}
  (\bibinfo{year}{2020}).

\bibitem[{\citenamefont{Pavanello et~al.}(2010)\citenamefont{Pavanello, Tung,
  and Adamowicz}}]{Pavanello:10}
\bibinfo{author}{\bibfnamefont{M.}~\bibnamefont{Pavanello}},
  \bibinfo{author}{\bibfnamefont{W.-C.} \bibnamefont{Tung}}, \bibnamefont{and}
  \bibinfo{author}{\bibfnamefont{L.}~\bibnamefont{Adamowicz}},
  \bibinfo{journal}{Phys. Rev. A} \textbf{\bibinfo{volume}{81}},
  \bibinfo{pages}{042526} (\bibinfo{year}{2010}).

\bibitem[{\citenamefont{Pyykk{\"o}}(2018)}]{Pyykko:18}
\bibinfo{author}{\bibfnamefont{P.}~\bibnamefont{Pyykk{\"o}}},
  \bibinfo{journal}{Mol. Phys.} \textbf{\bibinfo{volume}{116}},
  \bibinfo{pages}{1328} (\bibinfo{year}{2018}).

\bibitem[{\citenamefont{Ramsey}(1956)}]{Ramsey:56}
\bibinfo{author}{\bibfnamefont{N.}~\bibnamefont{Ramsey}},
  \emph{\bibinfo{title}{Molecular Beams}} (\bibinfo{publisher}{Oxford University
  Press, Oxford}, \bibinfo{year}{1956}).

\bibitem[{\citenamefont{Ramsey and Lewis}(1957)}]{Ramsey:57}
\bibinfo{author}{\bibfnamefont{N.~F.} \bibnamefont{Ramsey}} \bibnamefont{and}
  \bibinfo{author}{\bibfnamefont{H.~R.} \bibnamefont{Lewis}},
  \bibinfo{journal}{Phys. Rev.} \textbf{\bibinfo{volume}{108}},
  \bibinfo{pages}{1246} (\bibinfo{year}{1957}).

\bibitem[{\citenamefont{Brown and Carrington}(2003)}]{brown_carrington_2003}
\bibinfo{author}{\bibfnamefont{J.~M.} \bibnamefont{Brown}} \bibnamefont{and}
  \bibinfo{author}{\bibfnamefont{A.}~\bibnamefont{Carrington}},
  \emph{\bibinfo{title}{Rotational Spectroscopy of Diatomic Molecules}},
  Cambridge Molecular Science (\bibinfo{publisher}{Cambridge University Press, Cambridge, England},
  \bibinfo{year}{2003}).

\bibitem[{\citenamefont{{2018 CODATA recommended values}}(2018)}]{CODATA:18}
\bibinfo{author}{\bibnamefont{{2018 CODATA recommended values}}}
  (\bibinfo{year}{2018}),\url{https://physics.nist.gov/cuu/Constants}.

\bibitem[{\citenamefont{Komasa et~al.}(2020)\citenamefont{Komasa, Puchalski,
  and Pachucki}}]{Komasa:20}
\bibinfo{author}{\bibfnamefont{J.}~\bibnamefont{Komasa}},
  \bibinfo{author}{\bibfnamefont{M.}~\bibnamefont{Puchalski}},
  \bibnamefont{and} \bibinfo{author}{\bibfnamefont{K.}~\bibnamefont{Pachucki}},
  \bibinfo{journal}{Phys. Rev. A} \textbf{\bibinfo{volume}{102}},
  \bibinfo{pages}{012814} (\bibinfo{year}{2020}).

\bibitem[{\citenamefont{Harrick et~al.}(1953)\citenamefont{Harrick, Barnes,
  Bray, and Ramsey}}]{Harrick:53}
\bibinfo{author}{\bibfnamefont{N.~J.} \bibnamefont{Harrick}},
  \bibinfo{author}{\bibfnamefont{R.~G.} \bibnamefont{Barnes}},
  \bibinfo{author}{\bibfnamefont{P.~J.} \bibnamefont{Bray}}, \bibnamefont{and}
  \bibinfo{author}{\bibfnamefont{N.~F.} \bibnamefont{Ramsey}},
  \bibinfo{journal}{Phys. Rev.} \textbf{\bibinfo{volume}{90}},
  \bibinfo{pages}{260} (\bibinfo{year}{1953}).

\bibitem[{\citenamefont{Quinn et~al.}(1958)\citenamefont{Quinn, Baker,
  LaTourrette, and Ramsey}}]{Quinn:58}
\bibinfo{author}{\bibfnamefont{W.~E.} \bibnamefont{Quinn}},
  \bibinfo{author}{\bibfnamefont{J.~M.} \bibnamefont{Baker}},
  \bibinfo{author}{\bibfnamefont{J.~T.} \bibnamefont{LaTourrette}},
  \bibnamefont{and} \bibinfo{author}{\bibfnamefont{N.~F.}
  \bibnamefont{Ramsey}}, \bibinfo{journal}{Phys. Rev.}
  \textbf{\bibinfo{volume}{112}}, \bibinfo{pages}{1929} (\bibinfo{year}{1958}).

\bibitem[{\citenamefont{Code and Ramsey}(1971)}]{Code:71}
\bibinfo{author}{\bibfnamefont{R.~F.} \bibnamefont{Code}} \bibnamefont{and}
  \bibinfo{author}{\bibfnamefont{N.~F.} \bibnamefont{Ramsey}},
  \bibinfo{journal}{Phys. Rev. A} \textbf{\bibinfo{volume}{4}},
  \bibinfo{pages}{1945} (\bibinfo{year}{1971}).

\bibitem[{\citenamefont{Pachucki and Komasa}(2018{\natexlab{a}})}]{PK:18a}
\bibinfo{author}{\bibfnamefont{K.}~\bibnamefont{Pachucki}} \bibnamefont{and}
  \bibinfo{author}{\bibfnamefont{J.}~\bibnamefont{Komasa}},
  \bibinfo{journal}{Phys. Chem. Chem. Phys.} \textbf{\bibinfo{volume}{20}},
  \bibinfo{pages}{247} (\bibinfo{year}{2018}{\natexlab{a}}).

\bibitem[{\citenamefont{Pachucki and Komasa}(2018{\natexlab{b}})}]{PK:18b}
\bibinfo{author}{\bibfnamefont{K.}~\bibnamefont{Pachucki}} \bibnamefont{and}
  \bibinfo{author}{\bibfnamefont{J.}~\bibnamefont{Komasa}},
  \bibinfo{journal}{Phys. Chem. Chem. Phys.} \textbf{\bibinfo{volume}{20}},
  \bibinfo{pages}{26297} (\bibinfo{year}{2018}{\natexlab{b}}).

\bibitem[{\citenamefont{Pachucki and Komasa}(2019)}]{PK:19}
\bibinfo{author}{\bibfnamefont{K.}~\bibnamefont{Pachucki}} \bibnamefont{and}
  \bibinfo{author}{\bibfnamefont{J.}~\bibnamefont{Komasa}},
  \bibinfo{journal}{Phys. Chem. Chem. Phys.} \textbf{\bibinfo{volume}{21}},
  \bibinfo{pages}{10272} (\bibinfo{year}{2019}).

\bibitem[{\citenamefont{Bishop and Cheung}(1979)}]{Bishop:79}
\bibinfo{author}{\bibfnamefont{D.~M.} \bibnamefont{Bishop}} \bibnamefont{and}
  \bibinfo{author}{\bibfnamefont{L.~M.} \bibnamefont{Cheung}},
  \bibinfo{journal}{Phys. Rev. A} \textbf{\bibinfo{volume}{20}},
  \bibinfo{pages}{381} (\bibinfo{year}{1979}).

\bibitem[{\citenamefont{Ramsey}(1953)}]{Ramsey:53b}
\bibinfo{author}{\bibfnamefont{N.~F.} \bibnamefont{Ramsey}},
  \bibinfo{journal}{Phys. Rev.} \textbf{\bibinfo{volume}{89}},
  \bibinfo{pages}{527} (\bibinfo{year}{1953}).

\bibitem[{\citenamefont{Filin et~al.}(2020{\natexlab{b}})\citenamefont{Filin,
  Möller, Baru, Epelbaum, Krebs, and Reinert}}]{Filin:20a}
\bibinfo{author}{\bibfnamefont{A.~A.} \bibnamefont{Filin}},
  \bibinfo{author}{\bibfnamefont{D.}~\bibnamefont{Möller}},
  \bibinfo{author}{\bibfnamefont{V.}~\bibnamefont{Baru}},
  \bibinfo{author}{\bibfnamefont{E.}~\bibnamefont{Epelbaum}},
  \bibinfo{author}{\bibfnamefont{H.}~\bibnamefont{Krebs}}, \bibnamefont{and}
  \bibinfo{author}{\bibfnamefont{P.}~\bibnamefont{Reinert}},
  \bibinfo{journal}{arXiv}
  \eprint{2009.08911}.

\bibitem[{\citenamefont{Reid and Vaida}(1972)}]{Reid:72}
\bibinfo{author}{\bibfnamefont{R.~V.} \bibnamefont{Reid}} \bibnamefont{and}
  \bibinfo{author}{\bibfnamefont{M.~L.} \bibnamefont{Vaida}},
  \bibinfo{journal}{Phys. Rev. Lett.} \textbf{\bibinfo{volume}{29}},
  \bibinfo{pages}{494} (\bibinfo{year}{1972}).

\bibitem[{\citenamefont{Reid and Vaida}(1975)}]{Reid:75Errata}
\bibinfo{author}{\bibfnamefont{R.~V.} \bibnamefont{Reid}} \bibnamefont{and}
  \bibinfo{author}{\bibfnamefont{M.~L.} \bibnamefont{Vaida}},
  \bibinfo{journal}{Phys. Rev. Lett.} \textbf{\bibinfo{volume}{34}},
  \bibinfo{pages}{1064} (\bibinfo{year}{1975}).

\bibitem[{\citenamefont{Pohl et~al.}(2016)\citenamefont{Pohl, Nez, Fernandes,
  Amaro, Biraben, Cardoso, Covita, Dax, Dhawan, Diepold et~al.}}]{Pohl:16}
\bibinfo{author}{\bibfnamefont{R.}~\bibnamefont{Pohl}},
  \bibinfo{author}{\bibfnamefont{F.}~\bibnamefont{Nez}},
  \bibinfo{author}{\bibfnamefont{L.~M.~P.} \bibnamefont{Fernandes}},
  \bibinfo{author}{\bibfnamefont{F.~D.} \bibnamefont{Amaro}},
  \bibinfo{author}{\bibfnamefont{F.}~\bibnamefont{Biraben}},
  \bibinfo{author}{\bibfnamefont{J.~M.~R.} \bibnamefont{Cardoso}},
  \bibinfo{author}{\bibfnamefont{D.~S.} \bibnamefont{Covita}},
  \bibinfo{author}{\bibfnamefont{A.}~\bibnamefont{Dax}},
  \bibinfo{author}{\bibfnamefont{S.}~\bibnamefont{Dhawan}},
  \bibinfo{author}{\bibfnamefont{M.}~\bibnamefont{Diepold}},
  \bibnamefont{et~al.}, \bibinfo{journal}{Science}
  \textbf{\bibinfo{volume}{353}}, \bibinfo{pages}{669} (\bibinfo{year}{2016}).

\bibitem[{\citenamefont{Kalinowski et~al.}(2018)\citenamefont{Kalinowski,
  Pachucki, and Yerokhin}}]{Kalinowski:18}
\bibinfo{author}{\bibfnamefont{M.}~\bibnamefont{Kalinowski}},
  \bibinfo{author}{\bibfnamefont{K.}~\bibnamefont{Pachucki}}, \bibnamefont{and}
  \bibinfo{author}{\bibfnamefont{V.~A.} \bibnamefont{Yerokhin}},
  \bibinfo{journal}{Phys. Rev. A} \textbf{\bibinfo{volume}{98}},
  \bibinfo{pages}{062513} (\bibinfo{year}{2018}).

\end{thebibliography}
\end{document}